\documentclass[aps,prl,twocolumn,showpacs,amsmath,amssymb,superscriptaddress]{revtex4}
\usepackage{graphicx}
\usepackage{dcolumn}
\usepackage{bm}
\begin{document}
\title{\boldmath Scalar interaction limits from 
the $\beta$-$\nu$ correlation of trapped radioactive atoms}
\author{A.~Gorelov}
\affiliation{Department of Physics, Simon Fraser University, Burnaby, 
British Columbia, Canada V5A 1S6}
\author{D.~Melconian}
\affiliation{Department of Physics, Simon Fraser University, Burnaby, 
British Columbia, Canada V5A 1S6}
\author{W.P.~Alford}
\affiliation{University of Western Ontario, London, Ontario, 
Canada N6A 3K7}
\author{D.~Ashery} 
\affiliation{School of Physics and Astronomy, Tel Aviv University, 
69978 Tel Aviv, Israel}
\author{G. Ball}
\affiliation{TRIUMF, 4004 Wesbrook Mall, Vancouver, British Columbia, 
Canada V6T 2A3}
\author{J.A.~Behr}
\affiliation{TRIUMF, 4004 Wesbrook Mall, Vancouver, British Columbia, 
Canada V6T 2A3}
\author{P.G.~Bricault}
\affiliation{TRIUMF, 4004 Wesbrook Mall, Vancouver, British Columbia, 
Canada V6T 2A3}
{\author{J.M.~D'Auria}
\affiliation{Department of Chemistry, Simon Fraser University, Burnaby, 
British Columbia, Canada V5A 1S6}
\author{J.~Deutsch} 
\affiliation{Universit\'e Catholique de Louvain, B-1348 
Louvain-la-Neuve, Belgium}
\author{J.~Dilling} 
\affiliation{TRIUMF, 4004 Wesbrook Mall, Vancouver, British Columbia, 
Canada V6T 2A3}
\author{M.~Dombsky} 
\affiliation{TRIUMF, 4004 Wesbrook Mall, Vancouver, British Columbia, 
Canada V6T 2A3}
\author{P.~Dub\'e}
\affiliation{Department of Physics, Simon Fraser University, Burnaby, 
British Columbia, Canada V5A 1S6}
\author{J.~Fingler}
\affiliation{TRIUMF, 4004 Wesbrook Mall, Vancouver, British Columbia, 
Canada V6T 2A3}
\author{U.~Giesen}
\affiliation{TRIUMF, 4004 Wesbrook Mall, Vancouver, British Columbia, 
Canada V6T 2A3}
\author{F.~Gl\"uck}
\affiliation{KFKI RMKI, 1525 Budapest, POB 49, Hungary}
\author{S.~Gu}
\affiliation{TRIUMF, 4004 Wesbrook Mall, Vancouver, British Columbia, 
Canada V6T 2A3}
\author{O.~H\"ausser}
\altaffiliation{Deceased}
\affiliation{Department of Physics, Simon Fraser University, Burnaby, 
British Columbia, Canada V5A 1S6}
\author{K.P.~Jackson}
\affiliation{TRIUMF, 4004 Wesbrook Mall, Vancouver, British Columbia, 
Canada V6T 2A3}
\author{B.K.~Jennings}
\affiliation{TRIUMF, 4004 Wesbrook Mall, Vancouver, British Columbia, 
Canada V6T 2A3}
\author{M.R.~Pearson}
\affiliation{TRIUMF, 4004 Wesbrook Mall, Vancouver, British Columbia, 
Canada V6T 2A3}
\author{T.J.~Stocki}
\affiliation{Department of Physics, Simon Fraser University, Burnaby, 
British Columbia, Canada V5A 1S6}
\author{T.B.~Swanson} 
\affiliation{Department of Chemistry, Simon Fraser University, Burnaby, 
British Columbia, Canada V5A 1S6}
\author{M.~Trinczek}
\affiliation{Department of Chemistry, Simon Fraser University, 
Burnaby, British Columbia, Canada V5A 1S6}
\begin{abstract}
We have set limits on contributions of scalar interactions to nuclear $\beta$
decay. 
A magneto-optical trap (MOT) provides a localized source of atoms suspended in
space,
so the low-energy recoiling nuclei can freely escape and be detected
in coincidence with
the $\beta$. This allows reconstruction of the neutrino momentum, and the
measurement of the $\beta$-$\nu$ correlation, in a more direct fashion
than previously possible. The $\beta$-$\nu$ correlation parameter 
of the 
$0^+ \rightarrow 0^+$ pure Fermi decay of $^{38m}$K 
is $\tilde{a}$=0.9981$\pm$0.0030$^{+0.0032}_{-0.0037}$,
consistent with the Standard Model prediction $\tilde{a}$=1.
\end{abstract}
%Weak-interaction and lepton (including neutrino) aspects of nuclear physics
%Optical cooling of atoms; trapping
%Physics of Elementary particles;Other particles (including hypothetical)
%
\maketitle
%=============================================================================

%The angular distribution of neutrinos with respect to the beta direction in
%nuclear beta decay, i.e. the $\beta$-$\nu$ correlation,
The angular correlation of neutrinos and beta's in
nuclear beta decay
is historically one of the main experimental
probes of the vector and axial vector nature of the weak 
interaction~\cite{commins}.
%Accurate measurements of 
%the $\beta$-$\nu$ correlation in pure Fermi decay, which
%is directly sensitive to scalar interactions, 
%were not available until recently~\cite{garcia}.
A recent experiment using $^{32}$Ar decay is the only $\beta$-$\nu$
%correlation measurement in pure Fermi decay, which is directly
correlation measurement in pure Fermi decay, which is
sensitive to scalar interactions~\cite{garcia}.

%We deduce the $\beta^+$-$\nu$ correlation by measuring the $\beta^+$
%momentum and  the recoiling nucleus momentum
%in coincidence.
%We use a magneto-optical trap~\cite{raab} to provide a %cooled, 
We use a magneto-optical trap (MOT)~\cite{raab} to provide a %cooled, 
backing-free
source of atoms with well-defined position and negligible thermal energy.
%This allows us
We then
detect the low-energy nuclear recoil in coincidence with the emitted 
%$\beta^+$ event-by-event, and directly deduce the $\nu$ direction
$\beta^+$, and directly deduce the $\nu$ direction
and the $\beta^+$-$\nu$ correlation.
We also determine critical response functions of our 
detectors in situ from the decays themselves. 
Atom trap $\beta$-$\nu$ experiments are also pursued 
elsewhere~\cite{scielzoprl}.

In the 0$^+\rightarrow 0^+$ Fermi decays the leptons carry 
away no net angular momentum. Back-to-back $\beta$-$\nu$ emission is
forbidden in the Standard Model, because the W vector boson exchange produces
leptons with opposite helicity and their spins add to one.
%Back-to-back emission would be maximal for scalar boson exchange, 
%which produces
%leptons with the same helicity.  
%If we write the angular distribution
The angular distribution is
\begin{equation}
\nonumber  W (\theta_{\beta \nu}) = 1 + b \frac{m_{\beta}}{E_{\beta}} + 
a \frac{{\rm v}_{\beta}}{{\rm c}} 
{\rm cos}(\theta_{\beta \nu}).
\end{equation}
The $\beta$-$\nu$ 
coefficient 
$a$ is +1 for W exchange, 
and $a$ is --1 for a scalar boson producing same-helicity leptons.
In terms of scalar coupling constants $C_S$ and $C_S'$, and 
assuming for simplicity $C_V$=$C_V'$=1 ~\cite{JTW}: 
%\begin{equation}
\begin{eqnarray*}
%\begin{displaymath}
 a=[2-(|C_S|^2+|C_S'|^2)]/(2+|C_S|^2+|C_S'|^2), \\
 b=-2\sqrt{1-(\alpha Z)^2}Re(C_S + C_S')/(2+|C_S|^2+|C_S'|^2)
%\end{equation}
%eqnarray* has no equation numbers, or use \nonumber on each line
\end{eqnarray*}
%\end{displaymath}
The limit on the Fierz interference term $b$ from the 
%lack of 
dependence of
$0^+$$\rightarrow$$0^+$ decay strengths on $\langle E_{\beta} \rangle$ 
is very stringent, $b$=$-$0.0027$\pm$0.0029~\cite{towner},
but the coupling $C_S$+$C_S'$
describes
scalars that couple only to the left-handed $\nu$. 
Measurements of $a$ constrain scalar interactions independent of chirality or
time-reversal properties~\cite{garcia,herczeg}. 

The isobaric analog decays of the pure Fermi transitions
are well characterized.
Lowest recoil-order corrections 
to the allowed approximation value of 
$a$=1 
($<$3$\times$10$^{-4}$ in our case, $^{38m}$K)
do not depend on nuclear structure, and higher order corrections are
$<$0.0002~\cite{holstein,gluck}. 
Radiative corrections (see below)~\cite{gluck}
can also be calculated to the order required 
independent of nuclear structure.
In addition, $^{38m}$K decay is known to proceed cleanly to the 
ground state, with experimental limits on excited-state branches of 
$<$ 2$\times$$10^{-5}$ \cite{hagberg}. 
Disagreement with $a$=1 greater than these corrections 
would be from a standard model-violating scalar
interaction.  

A scalar term could be
produced by the exchange of scalar bosons found in 
standard model extensions~\cite{herczeg}. 
There is a phenomenological window open for sleptons, 
although couplings to the first generation are thought to be 
small~\cite{herczeg}. 
A QCD-induced scalar interaction 
is a second-class current that is $<$5$\times$10$^{-5}$ in the Standard Model~\cite{holsteinscalar}. 
$\beta$-$\nu$ correlations are complementary to 
scalar constraints from 
pseudoscalar $\pi$$\rightarrow$$\nu$$e$ decay~\cite{campbell}. 
The possibility of constraining scalar interactions from loop corrections 
%to the neutrino mass matrix is considered in Ref.~\cite{ito}. 
to $\nu$ masses is considered in Ref.~\cite{ito}. 
%The best previous $\beta$-$\nu$ experiment used the 
%$\beta$-delayed 
%protons from the Fermi decay of 
%$^{32}$Ar to determine $\tilde{a}$=$a/(1+b\frac{m_{\beta}}{\langle E_{\beta}\rangle})$ 
The best previous $\beta$-$\nu$ experiment used 
$\beta$-delayed 
protons from  
$^{32}$Ar decay to determine $\tilde{a}$=$a/(1+b m_{\beta}/\langle E_{\beta}\rangle)$ 
to be 0.9989$\pm$0.0052$\pm$0.0036~\cite{garcia}.

Our apparatus was also used to constrain
massive $\nu_x$-$\nu_e$ admixtures~\cite{trinczek}.
%A mass-separated ion beam of $^{38m}$K 
A beam of $^{38m}$K ions
($t_{1/2}= 0.924$~s, $Q_{\beta^+}=5.022$~MeV) is
produced at TRIUMF's ISAC 
%and TISOL
facilities~\cite{isacref},
stopped and released as neutral atoms with a $900^\circ$C Zr 
foil~\cite{melcon}, and captured with 
%stopped and released as neutral atoms~\cite{melcon}, 
%and captured with 
$\approx10^{-3}$ efficiency in a vapor-cell
%you put ``magneto-optical trap'' here, we have moved (MOT) definition to
%the first time ``magneto-optical trap'' appears in 2nd paragraph, p. 1 
MOT.
%\@~\cite{gorelov}.  
The MOT traps only the 
$^{38m}$K and none of the ground state $^{38}$K\ beam contaminant. 
To escape backgrounds from 
untrapped atoms
of %both 
$^{38}$K and 
$^{38m}$K, 
we transfer the trapped atoms 
with 75\% efficiency by a chopped laser push beam
to a 2$^{\mathrm{nd}}$ MOT equipped with the nuclear 
detectors (Fig.~\ref{fig1}).
%The duty cycle entails the following:
The duty cycle entails:
push atoms from the first trap for 20 ms; wait 50 ms to
transfer; 
change the 2$^{\rm nd}$ MOT laser frequency and power to minimize cloud size; 
wait 1 ms to let cloud reach equilibrium; 
count for 150 ms; repeat~\cite{swanson}. %??? CHECK DUTY CYCLE 
No atoms are lost
from the trap during the frequency switch. 
%After transfer, the 
%frequency and power of the 2$^{\mathrm{nd}}$ MOT beams are 
%changed to minimize the cloud size.
%The MOT force on the K atoms consists of near-resonant
The MOT force uses
laser light and a weak (dB$_z$/dz = 20~G/cm) magnetic 
quadrupole field, so the Ar
recoils 
escape the trap without perturbation.
We accelerate the positive Ar ions
produced by electron shakeoff~\cite{carlsonhe6}
%~\cite{gorelov,carlson}
with a uniform electrostatic field 
to separate them in time-of-flight (TOF) 
from the neutral Ar$^{0}$ 
atoms. 
%, and collect most of their angular distribution.

\begin{figure}[t]
  \centering\includegraphics[width=3in,angle=0]{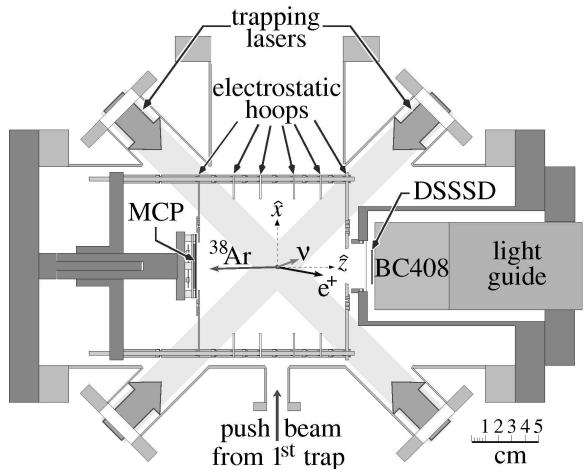}
  \caption{Top view of the 2$^{\mathrm{nd}}$ MOT apparatus with 
    the recoil and $\beta$ detectors.
%    the recoil (MCP) and $\beta$ (DSSSD and BC408 scintillator) 
%    detectors.
    \label{fig1}}
\end{figure}

The $\beta$-telescope is a 
position-sensitive 
22x22x0.49~mm double-sided 
Si-strip detector (DSSSD) 
backed by a 
%with energy resolution 7.9\% at 1.6~MeV\@.
%We exclude the 1 mm outer strips in X and Y from the analysis, as 
%the collection efficiency for charge there and from Si regions 
%outside the strip area is difficult to model and measure.  
%The solid angle of the beta telescope is 1\%.
%$\varnothing 6.5~\mathrm{cm}\times5.5$~cm long BC408 plastic 
%scintillator. 
$\varnothing 6.5~\mathrm{cm}\times5.5$~cm long BC408 plastic 
scintillator. 
The telescope is separated from the trap vacuum by a 125$\mu$m thick
%plastic scintillator, and is 
%separated from the trap vacuum by a 125$\mu$m thick
Be foil located 2mm from the DSSSD to minimize angle straggling.
The telescope coincidence rejects 95\% of the 2.17 MeV $\gamma$-rays from
untrapped $^{38}$K ground state. 
The gain is actively stabilized
at the low count rates of $<$ 200 Hz 
using a stabilized light pulser.
%Please consider replacing ``light'' with ``LED'', but not if you insist on
%writing it out. 
%we think it loses content to not use LED, but don't have room to write it out
%by 
%a transistorized actively stabilized photobase and stabilized light pulser.
%you have changed to something like
%and light emitting diode (LED) or photodiode active stabilization
%and LED/photodiode active stabilization. 

The Ar recoils, which have 0--430~eV of initial kinetic energy, 
are detected by a Z-stack of three uncoated microchannel plates (MCP).
A fixed aperture defines a 24.0~mm active diameter for the 
TOF[E$_{\beta}$] analysis (see below).
The resistive anode position readout 
is calibrated with a mask and 
an $\alpha$-source to have 0.25~mm resolution within 
the 20~mm diameter
used for the reconstructed angular distribution 
analysis.  
The $E$ field  
accelerates the Ar$^{+1}$ ions 
to 4.8--5.3 keV.
We measured the  
MCP efficiency in this energy 
range to be constant to accuracy 0.0060
by 
comparing the rate of $\beta$-recoil
coincidences for four values of $E_{\hat z}$.
The $\beta$-$\nu$ correlation analysis is
done with the ions, because 
the efficiency for neutral recoils is not as well understood.
The angle dependence of the MCP efficiency was assumed constant over
the small impact angles of $\pm$$5^{\rm o}$,
with error (Table~\ref{tab-1}) 
spanning the small effect seen in the literature~\cite{fraser},
consistent with our analysis of 
recoils that uniformly illuminate
the MCP.

We maintain a population of $\approx$ 2,000 atoms of $^{38m}$K
in the detection MOT.
The trap lifetime, limited by residual gas, is 45~s,
so 97\% of the $^{38m}$K atoms decay while in the trap.
%There is a $\beta^+$ singles background of $<$ 2\% and a 
%negligible coincidence background from atoms on the walls. 
%We measured this by releasing the atoms to the walls, and the measured
%coincidence backgrounds were consistent with accidental coincidences in
%the detectors.  
Atoms on the walls produce  
a $\beta^+$ singles background of $<$ 2\% and a 
negligible coincidence background consistent with accidental coincidences,
measured by deliberately releasing the atoms to the walls.
%We measured this by releasing the atoms to the walls, and the measured
%coincidence backgrounds were consistent with accidental coincidences in
%the detectors.  
Ions from the walls are excluded from the MCP by the electric field. 
Ions from the trap strike no material before reaching the MCP.
The $E$ field electrodes are made from glassy carbon to minimize $\beta^+$ 
scattering effects.

The 
average 
trap-MCP distance is determined to be $61.08\pm0.01$~mm 
from a
fit to the leading edge of the TOF peak of the fastest Ar$^{0}$
recoils (Fig.~\ref{fig2}). 
These were shown to be Ar$^0$ by
%what you wrote is incorrect, the field is not varied continuously, so we
%have rephrased:
applying E fields of 400 and 800 V/cm;
%using E fields of 400 and 800 V/cm; 
%varying the E field from 400 to 800 V/cm; 
the leading edge was undistorted by any
detection  
%of the known $\tau_e=260$~ns~\cite{arminusexp} Ar$^-$
of the $\tau_e=260$~ns~\cite{arminusexp} Ar$^-$
metastable state.
%any 
%distortion of the leading edge
%from production  
%of the known $\tau_e=260$~ns~\cite{arminusexp} Ar$^-$
%metastable state was unmeasurably small. 

We image the cloud by photoionizing a small fraction 
of the $^{38m}$K atoms with a pulsed laser and accelerating
them to the MCP. The 
$\hat{x}$, $\hat{y}$, and $\hat{z}$ distributions (see Fig. 1) are fit
well with Gaussians of 0.8, 1.1, and 0.65~mm FWHM.
The $\hat{z}$ distribution (along the trap-MCP axis) 
limits the timing resolution for
Ar$^{+1}$ recoils to 5 ns.
Two CCD cameras 
image the trap laser fluorescence, and the trap centroid was kept constant
to $\pm0.05$~mm.

%A 2D scatter plot of the ion data is shown in 
\begin{figure}[t]
%  \centering\includegraphics[width=3.1in]{trinczek-fig2.ps}
%  \centering\includegraphics[width=3in,angle=0]{gorelov-fig2-smaller2.eps}
%  \centering\includegraphics[width=3in,angle=0]{fig2_scalpaper_blue-red.eps}
%  \centering\includegraphics[width=3in,angle=0]{fig2_scalpaper_blue-red_gimp2.eps}
%  \centering\includegraphics[width=3in,angle=0]{fig2_scalpaper_red_gimpiaeg200.eps}
%  \centering\includegraphics[width=3in,angle=0]{fig2_scalpaper_red_filt.eps}
  \centering\includegraphics[width=3in,angle=0]{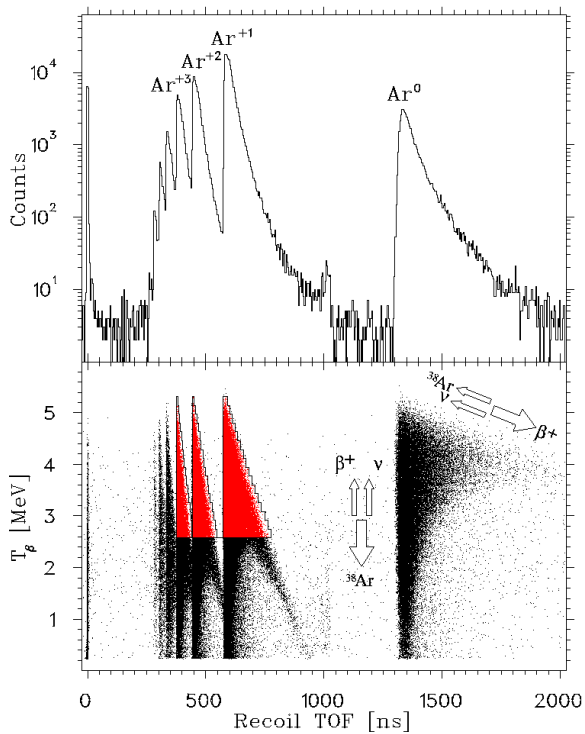}
  \caption{(color online)
Bottom:    Scatter plot of recoil TOF vs. $T_{\beta}$ with one dot shown for
each of 500\,000 events. The suppressed 
back-to-back lepton emission produces longer recoil TOF.
The $E$ field separates the Ar charge states. The analysis cuts are shown.
Top: TOF projections of the 2D scatter plot. The 0.1\% background at 
TOF$\approx$1020 ns is from $\beta$'s scattering off the 
MCP into the $\beta$ telescope and
can be rejected kinematically.
    \label{fig2}}
\end{figure}

We performed two independent analyses of the data set.
%In the first analysis, we fit the experimental TOF spectra of ion
In the first analysis, we fit the TOF spectra of ion
recoils for various $\beta^+$ energy cuts (Fig.~\ref{fig3})
to a Monte Carlo (MC) simulation 
based on GEANT~\cite{geant}. 
%Qualitatively, for fixed E$_\beta^+$, the recoil TOF 
Qualitatively, for fixed E$_\beta$, the recoil TOF 
increases monotonically with cos($\theta_{\beta \nu})$. The TOF of the
ions with most sensitivity to $a$ increases with decreasing $E_{\beta}$.
%We fit simultaneously to Ar$^{+1}$, Ar$^{+2}$ and Ar$^{+3}$ charge states.
We fit simultaneously to Ar$^{+1,+2,+3}$ charge states.

In the second 
analysis we use the complete momentum information measured for the
$\beta$ and the recoil to deduce the momentum of the $\nu$ and the 
$\beta-\nu$ angle
(Fig. \ref{fig3}). 
The kinematics are over-determined for 
p$_{\rm recoil}$$<$ $Q_{\beta+}$/c~\cite{kh}, 
so this was done either using the measured $E_{\beta}$
or determining it from the recoil momentum. 
The measured
%angular distribution is fit to the MC as a function of $a$, and
angular distribution is fit to the MC simulation as a function of $a$, and
agrees with the TOF[E$_{\beta}$]
analysis.

\begin{figure}[tbh]
%  \centering\includegraphics[width=3.1in,angle=0]{justalexandshake4.eps}
%  \centering\includegraphics[width=3.1in,angle=0]{fig3_scalpaper.eps}
%   \centering\includegraphics[width=3.1in,angle=0]{gorelov-fig3.eps}
%   \centering\includegraphics[width=3.1in,angle=0]{fig3_scalpaperjan3.eps}
   \centering\includegraphics[width=3.1in,angle=0]{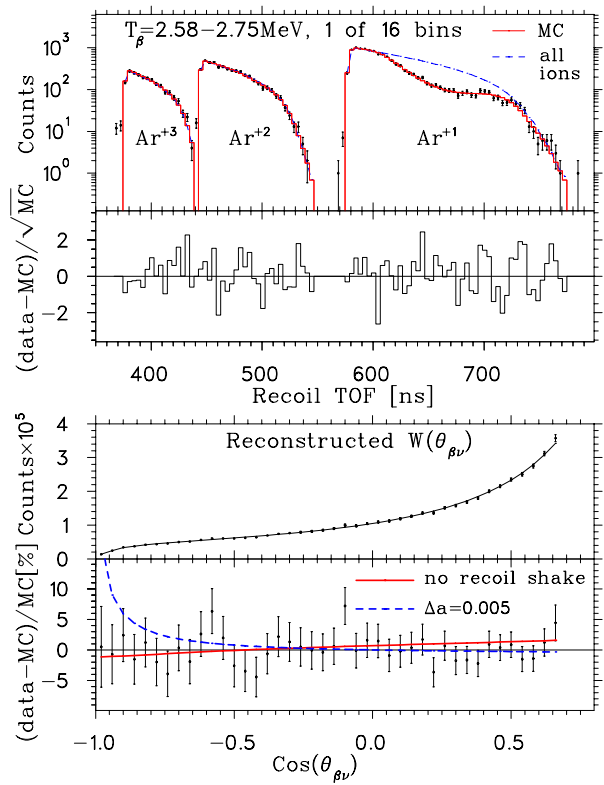}
  \caption{(color online)
    Top: lowest of 16 $T_{\beta}$ bins for the MC fit to  
the Ar$^{+1,+2,+3}$ data, 
and residuals. The confidence level for the entire fit is 52\%.
Data has been binned to show sensitivity to $a$. 
Fits are done with 4 ns bins. The dip in Ar$^{+1}$ is from the finite MCP
size: the dashed curve has an artificially larger MCP collecting all ions. %??? 
%Changes of $a$ of 1\% produce 10\% effects at long TOF.
 Bottom: Fit to reconstructed angular distribution, showing residuals.
%This would be a straight line if all 
%ions were collected.  
Also shown are the effect of a change in $a$ by 0.005, 
and the effect of recoil energy-dependent 
shakeoff. 
    \label{fig3}}
\end{figure}

We present
the details of the TOF[E$_{\beta}$] 
analysis and detailed evaluation of systematic errors. This 
analysis 
lets us constrain
critical physical and instrumental effects, 
but requires excellent $\beta$ telescope 
characterization.
Table~\ref{tab-1} shows systematic errors, determined by MC simulations
varying each parameter by its possible error and determining its effect
on $a$, with other parameters refit as appropriate.
Some errors in the table are 
summaries of more than one correlated systematic error. 
These line items are
uncorrelated, so we add them in quadrature to determine the
total systematic error.

\begin{table}[htb]
\begin{tabular*}{3.4in}{ll}
$\langle  E \rangle $ field/trap width: &  0.0017 \\
$E$ field nonuniformity & 0.0014 \\
E$_{\beta +}$ Detector Response: & \\
\hspace*{4mm} \indent Lineshape tail/total 0.101$\pm$0.006 &  0.0006 \\
%\hspace*{4mm} \indent 511keV Compton summing/total 0.068$\pm$0.004  & 0.0009\\
\hspace*{4mm} \indent 511keV Compton summing/total to 10\%  & 0.0009\\
\hspace*{4mm} \indent  Calibration including nonlinearity  & 0.0017\\ 
MCP Eff[E$_{\rm Ar+}$] measured constant 4.8-5.3keV 
 & $0.0007$\\
MCP Eff[$\theta$]/XY trap position  &0.0008\\ 
e$^-$ shakeoff dependence on p$_{\rm recoil}$ $s=0^{-0}_{+.014}$ 
& ~$^{+0}_{-0.0018}$\\ % 
\end{tabular*}
\caption{List of $\tilde{a}$ uncorrelated systematic errors. 
%There is no
%correction to $\tilde{a}$ from any of these errors.
}%
%, except e$^-$ shakeoff (see text).}
\label{tab-1}
\end{table}

We can test the MC simulation of the 
%$\beta^+$ energy lineshape response 
%with the $\beta$-recoil coincidences, because $E_{\beta}$
%is uniquely defined
%by the recoil TOF
%for p$_{\rm recoil}$ $<$ $Q_{\beta+}$/c ~\cite{kh}. 
$\beta^+$ energy lineshape  
with the $\beta$-recoil coincidences.
From Fig.~\ref{fig2},
the $E_{\beta}$ spectra in coincidence with neutral recoils for  
TOF intervals from 1500 to 1800 ns are peaks determined by the detector 
resolution and the angular 
acceptance, and a tail determined by $\beta^+$ backscatter, bremsstrahlung,
and $\sim$20\% of the tail from scattering off inactive volumes. 
From this and from detailed
kinematic reconstruction of the $\beta$-Ar$^{+1}$ coincidences, 
we have determined that both the size of this tail and the  
511 keV Compton summing agree with the MC simulation; they are listed in Table I.

%Please remove this ``simulation''. MC here is an adjective modifying fit.
The E$_{\beta}$ calibration is determined by a MC fit to the energy spectrum 
in coincidence with recoils with 370 ns $\leq$ TOF $\leq$ 900 ns, which 
includes all the observed Ar$^{+1,+2,+3}$ recoils.
We use the expression $x_{ADC}= x_{0}$ + $c_2$  T$_{\beta}$/(1+$qT_{\beta}$),
with nonlinear term $q=$(0.33$\pm$1.49)$\times$10$^{-3}$ MeV$^{-1}$.
The calibration parameters are not sensitive to the value of $a$. 
%Use of an energy calibration determined from the 
Use of an E$_{\beta}$ calibration determined from the 
$\beta$ singles spectrum over the fit range  
produces a value of $a$ consistent within the error in Table I.
The fit to the coincidence energy spectrum has 
$\chi^2$/N=21.8/23, and the fit to the $\beta$ singles energy 
spectrum has $\chi^2$/N=10.8/11.
We use the Fermi function and corrections of ~\cite{wilkinson}.  

We are working to extend the experiment to 
%$T_{\beta}\leq$2.5 MeV,
%with the goal to independently determine $b$. The quality of the fits becomes 
%considerably poorer, and as yet the systematic errors from $\beta$ scattering, 
$T_{\beta}\leq$2.5 MeV
to independently determine $b$. The fits become
poorer, and systematic errors from $\beta$ scattering, 
the low-energy lineshape tail, and possible 
additional sources are not fully understood.
The $T_{\beta}$ cutoff eliminates almost all backscattered events, as well as
all possible contamination from untrapped $^{38}$K ground state decays.

We include order-$\alpha$ 
radiative corrections. These are dominated by undetected momentum
carried away by real bremsstrahlung
photons, which we include in the MC event generator~\cite{gluck2}.
They change $a$ by 0.003 in the $^{32}$Ar 
experiment~\cite{gluck,garcia}. 
Because the $\beta^+$ 
energy spectrum is also affected, and we use 
the $\beta^+$ 
spectrum itself for our
%energy calibration, our net result is that 
%the radiative 
%corrections change $a$ by considerably less in our experimental geometry.
E$_{\beta}$ calibration, our net result is that 
the radiative 
corrections change $a$ by considerably less in our experiment.

Three independent measures determine the 
$E$ field. 
The 
leading edge TOF of the Ar ion spectra implies
$E_{\hat z}$=807.7$\pm$0.16 V/cm, 
independent of
$a$ and consistent for all charge states.
The field nonuniformity is constrained  
by the TOF of the photoionized
$^{38m}$K atoms, and by 
a population of `wrong-way' recoils produced from $\beta$'s firing the MCP,
which give central values 807.7 and 
808.3 V/cm.
The nonuniformity is $<$ 1.0 V/cm/cm  %1.5
and the resulting
error in $a$ is 0.0014. % 0.0020 

We collect 89\%, 99.6\%, and 99.9\%  of the Ar$^{+1,+2,+3}$ %???
ions in coincidence for $T_{\beta}>2.58$ MeV. 
%with $\beta^+$'s in the fit region.
%, and
%higher percentages of higher charge states. 
This finite acceptance is a source
of systematic error (see Fig.~3); since a fixed aperture 
defines the MCP size, the 
acceptance contributes part of the dependence of $a$ on the $E$
field and trap position, as quantified by the MC analysis (Table I).
%As mentioned above, a fixed diameter 
%aperture defines the MCP acceptance, %for the TOF[E$_{\beta}$] analysis, 
%and the trap position is well-determined in all three dimensions.
%The finite acceptance is the largest reason for the dependence of
%$a$ on the $E$ field and the XY trap position, and these errors are
%defined by the detailed MC analysis. 

Dependence of the probability of electron shakeoff on the recoil ion 
energy has been seen in $^{6}$He $\beta^-$ decay~\cite{carlsonhe6}. 
A recent simple estimate relates this effect to oscillator strengths
and suggests that it is larger in 
$\beta^+$ decay~\cite{scielzo}. The recoil
energy spectrum to lowest order is distorted by (1+$s$E$_{\rm rec}$). 
We constrain this effect experimentally by fitting $s$ and $a$ simultaneously 
in our TOF[E$_{\beta}$] fit for Ar$^{+1,+2,+3}$. We only include $s$ 
in the Ar$^{+1}$ spectrum, because the model
of Ref.~\cite{scielzo} using semiempirical oscillator strengths~\cite{verner}
suggests that $s$ for Ar$^{+2}$ (or Ar$^{+3}$) would be 0.11 (or 0.05) 
the size of $s$ for Ar$^{+1}$.
%-0.0133 +- 0.0144 , upper limit 0.0195
We find $s$=$-$0.013$\pm$0.020, a result in a nonphysical region 
with one $\sigma$ upper limit $s < 0.014$ and 
%error on $a$ 
change in $a$: 
$\Delta_a=0^{+0}_{-0.0018}$. 
The estimate of 
Ref.~\cite{scielzo} is $s$=0.031. 
A similar fit of the Ar$^{+1}$
reconstructed angular distribution to $a$ and $s$ gives consistent result and 
error (Fig.~\ref{fig3}).
We can constrain $s$ and $a$ simultaneously
because the greatest sensitivity to $a$ is at the null in
the angular distribution, 
and because we fit as a function of E$_{\beta}$.
%We find $s$=0.008$\pm$0.021 (consistent with the estimate of 0.031 of
%Ref.~\cite{scielzo}), 
%producing a
%change in $a$ of 0.0002$\pm$0.0020. 
%A similar fit of the Ar$^{+1}$
%reconstructed angular distribution to $a$ and $s$ gives consistent result and 
%error (Fig.~\ref{fig3}).
%We can constrain $s$ and $a$ simultaneously
%because the greatest sensitivity to $a$ is at the null in
%the angular distribution, 
%and because we fit as a function of E$_{\beta}$.
A fit to the total TOF spectrum summed over all E$_{\beta}$ 
would be more strongly correlated with the recoil momentum spectrum.

Our fit values for  
$a$ and $b$ are strongly correlated 
in the $E_{\beta}$ region used. 
Although we fit as a function of $E_{\beta}$, careful investigation of the 
correlations shows that the physical observable we report here is, for
$|b|$$<$0.04, effectively 
indistinguishable from that reported by Ref.~\cite{garcia}, 
$\tilde{a}$= $a/(1+bm_{\beta}/ \langle E_{\beta} \rangle)$,  
but with $\langle E_{\beta} \rangle$=3.3 MeV.
We find
%$\tilde{a} = 0.9981\pm0.0030\pm0.0037$, in agreement with the Standard Model. 
$\tilde{a} = 0.9981\pm0.0030^{+0.0032}_{-0.0037}$, 
in agreement with the Standard Model. 
If we vary $b$ from $-$0.0075 to +0.0021, 
the 90\% confidence range 
of the Fierz interference term limits in Ref.~\cite{towner}, 
then $a$ changes from 0.9971 to 0.9984, while $\tilde{a}$ changes
by $<$1$\times$10$^{-4}$.
%only 9$\times$10$^{-5}$ to -3$\times$10$^{-5}$.   
%by only 9.2$\times$10$^{-5}$ to -2.6$\times$10$^{-5}$.   
%atilde changes by about +8.0x10**-5 and -2.3x10**-5
Our measurement has comparable errors to
Ref.~\cite{garcia} with an entirely different
experimental method. 

%We dedicate this work to the memory of O.~H\"ausser. 
We acknowledge TRIUMF/ISAC staff
and comments by A.R. Young, 
R.M. Woloshyn, J. Ng, B.A. Campbell, I.A. Towner,
and
P. Herczeg. 
Supported by the National Research Council 
of Canada 
through TRIUMF, the Natural Sciences and Engineering Research Council 
of Canada, the Israel Science Foundation, and WestGrid.

%

%=============================================================================
\end{document}